\documentclass[twocolumn,pre,superscriptaddress,nofootinbib]{revtex4}

\usepackage{dcolumn}
\usepackage{amsmath}

\usepackage{graphicx}

\newcommand{\half}{\tfrac12}
\newcommand{\etal}{{\it{}et~al.}}
\newcommand{\defn}{\textit}
\newcommand{\Ord}{\mathrm{O}}
\newcommand{\set}[1]{\lbrace#1\rbrace}

\begin{document}
\title{Improved mutual information measure for classification and community detection}

\author{M. E. J. Newman}
\affiliation{Department of Physics, University of Michigan, Ann Arbor,
  Michigan, USA}
\affiliation{Center for the Study of Complex Systems, University of
Michigan, Ann Arbor, Michigan, USA}
\author{George T. Cantwell}
\affiliation{Department of Physics, University of Michigan, Ann Arbor,
  Michigan, USA}
\author{Jean Gabriel Young}
\affiliation{Center for the Study of Complex Systems, University of
Michigan, Ann Arbor, Michigan, USA}

\begin{abstract}
The information theoretic quantity known as mutual information finds wide use in classification and community detection analyses to compare two classifications of the same set of objects into groups.  In the context of classification algorithms, for instance, it is often used to compare discovered classes to known ground truth and hence to quantify algorithm performance.  Here we argue that the standard mutual information, as commonly defined, omits a crucial term which can become large under real-world conditions, producing results that can be substantially in error.  We demonstrate how to correct this error and define a mutual information that works in all cases.  We discuss practical implementation of the new measure and give some example applications.
\end{abstract}

\maketitle

\section{Introduction}
Mutual information is widely used in physics, statistics, and machine learning as a tool for comparing different labelings of a set of objects~\cite{CT91}.  For instance, within physics it is used in statistical mechanics for comparing states of spin models~\cite{WTV11} and particularly in network science for comparing partitions of networks into communities, mutual information being perhaps the standard measure for quantifying the performance of community detection algorithms~\cite{DDDA05,Fortunato10}: it tells us the extent to which the set of communities found by an algorithm agree with a given set of ground-truth communities.  In machine learning and statistics, mutual information is similarly used in classification problems to quantify the similarity of different labelings of sets of objects~\cite{AR14}.  For instance, we might attempt to deduce characteristics of a set of users of an online service, such as their age group or gender, and then calibrate our algorithm by using mutual information to compare our results against known characteristics of a test set of users.

Imagine then that we have some set of individuals or objects, such as people, documents, email messages, or aerial photographs, among many other possibilities.  Each object can be classified or labeled as belonging to one of several types, groups, or communities.  People could be labeled by sex, race, or blood type for instance; documents by topic; aerial photographs by type of terrain, and so forth.  Now imagine we have two different sets of labels for our objects, one inferred by some algorithm and the other assigned for instance by human experts.  The mutual information of the two labelings represents the amount of information that the first labeling gives us about the second---in effect, how good the algorithm is at mimicking the human experts.  If the two labelings are exactly the same, then the first tells us everything about about the second and the mutual information is maximal.  If the two labelings are completely unrelated then the first tells us nothing about the second and the mutual information is zero.  A particularly nice feature of mutual information is that it is invariant under permutations of the labels of one or both labelings, meaning that one does not have to worry about whether the labels match up.

Despite its advantages, however, mutual information is not perfect.  In this paper we show that, as traditionally defined, the mutual information can give inaccurate answers---indeed maximally inaccurate---under certain conditions and particularly when the number of groups or communities differs between the two labelings.  In some applications these conditions arise frequently, in which case the standard measure can fail badly.  We show how to properly correct for this failure, creating a mutual information measure that gives useful answers under all circumstances.

Various shortcomings of the mutual information have been discussed previously and several alternatives or variants have been proposed.  Best known is the work of Danon~\etal~\cite{DDDA05}, who point out that the range of values of the mutual information is somewhat arbitrary and suggest a modified measure, the \defn{normalized mutual information}, whose values run from zero to one.  This measure is useful, for instance, for comparing results across data sets of different sizes and has become widely used, particularly in network science.  We discuss the normalized mutual information in more detail in Section~\ref{sec:mi} and define a normalized version of our own measure in Section~\ref{sec:shared}.  

Several authors have raised objections to the mutual information similar to those discussed in this paper and have suggested potential solutions~\cite{Meila07,RH07,Dom02,AGAV09}.  Best known of these is the \defn{variation of information}, proposed by Meil\u{a}~\cite{Meila07}, a metric information distance that sees some use with clustering and community detection algorithms.  Though formally elegant, the variation of information is difficult to interpret, bearing no simple relationship to the labelings it describes.  Rosenberg and Hirschberg~\cite{RH07} have proposed \defn{V-measure}, a heuristic one-parameter family of measures that emphasizes a balance between two desirable properties, homogeneity and completeness.  Perhaps closest to our own work is that of Dom~\cite{Dom02}, who proposes a measure that shares some features with ours but which, as we will argue, has problems of its own.  A useful survey of these and other measures has been given by Amig\'o~\etal~\cite{AGAV09}.

A different shortcoming of the mutual information has been highlighted by Vinh~\etal~\cite{VEB10}, who argue that the mutual information should by rights be zero for unrelated random labelings, but in practice this is only true in the limit of infinitely large sets of objects.  For finite sets, and hence for all practical applications, the expected mutual information of two random labelings is nonzero, which is undesirable.  Vinh~\etal\ propose a measure they call the \defn{adjusted mutual information}, which corrects for this problem by subtracting the average mutual information of a random pair of labelings.  We will show that our own proposed measure also corrects for this shortcoming of the mutual information, but we argue that it does so in a more correct manner.  A detailed discussion and comparison of the various measures is given in Section~\ref{sec:discussion}.

\section{Mutual information and conditional entropy}
\label{sec:mi}
Consider a set of $n$ objects, numbered 1~to~$n$, and consider two divisions of those objects into groups or communities.  The two divisions need not have the same number of groups but the groups must be non-overlapping within each division and they must not be empty.  Let us suppose that the first division has $R$ groups labeled by integers~$r=1\ldots R$ and the second division has $S$ groups labeled $s=1\ldots S$.  Let $a_r$ be the number of objects with label~$r$ in the first labeling, let~$b_s$ be the number with label~$s$ in the second, and let $c_{rs}$ be the number labeled~$r$ in the first labeling and $s$ in the second.  Note that
\begin{align}
\label{eq:sumrules1}
&a_r = \sum_{s=1}^S c_{rs}, \qquad
b_s = \sum_{r=1}^R c_{rs}, \\
\label{eq:sumrules2}
&\sum_{r=1}^R a_r = \sum_{s=1}^S b_s = \sum_{rs} c_{rs} = n,
\end{align}
and that $a_r>0$ and $b_s>0$ for all $r$ and $s$ since the groups are non-empty.  With these definitions we can compute the probabilities~$P(r)$ and~$P(s)$ that an object chosen uniformly at random has group label~$r$ or $s$, or the probability~$P(r,s)$ that it has both labels~$r$ and~$s$, from
\begin{equation}
P(r) = {a_r\over n}, \qquad
P(s) = {b_s\over n}, \qquad
P(r,s) = {c_{rs}\over n}.
\label{eq:prs}
\end{equation}

Now we ask the following question: if we are told the label~$r$ of a particular object, how much additional information is need to specify the other label~$s$ of the same object, on average?  Information theory tells us that the answer is given by the entropy of the conditional probability distribution~$P(s|r) = P(r,s)/P(r)$:
\begin{equation}
\mbox{Entropy} = -\sum_s P(s|r) \log P(s|r).
\end{equation}
The average of this entropy over all objects, i.e.,~over the complete distribution of~$r$, is the quantity we call the \defn{conditional entropy} of $s$ given~$r$:
\begin{align}
H(s|r) &= -\sum_r P(r) \sum_s P(s|r) \log P(s|r) \nonumber\\
  &= -\sum_{rs} P(r,s) \log {P(r,s)\over P(r)}.
\label{eq:conditional1}
\end{align}
The conditional entropy is the average amount of additional information we would need to supply, on top of the value~$r$, in order to specify the value~$s$.  (Traditionally the logarithms would be taken base~2, giving entropy in units of bits, but other choices are possible and produce only an overall multiplier in the entropy.  None of our results will depend on what base is used.  Note also that for the expression above and subsequent expressions to give correct answers we must adopt the convention that $0 \log 0 = 0$.)

The maximum of the conditional entropy occurs when $r$ and $s$ are independent, so that $r$ tells us nothing about~$s$ and $P(r,s) = P(r) P(s)$, which gives $H(s|r) = H(s)$ where
\begin{equation}
H(s) = -\sum_s P(s) \log P(s)
\label{eq:entropy}
\end{equation}
is the (unconditional) entropy of~$s$.  Conversely, the minimum value of the conditional entropy is zero, which occurs when $r$ and $s$ are in perfect agreement.

Arguably the conditional entropy is a little counter\-intuitive since it is minimized, not maximized, when labelings are identical.  One can reverse the scale and create a measure that is maximized for maximum similarity by subtracting $H(s|r)$ from its maximum value~$H(s)$, which gives the quantity known as the \defn{mutual information}:
\begin{align}
I(r;s) &= H(s) - H(s|r) \nonumber\\
  &= - \sum_s P(s) \log P(s) + \sum_{rs} P(r,s) \log {P(r,s)\over P(r)}
     \nonumber\\
  &= \sum_{rs} P(r,s) \log {P(r,s)\over P(r)P(s)}.
\label{eq:mi}
\end{align}
This measure is now zero for uncorrelated labels and equal to $H(r)=H(s)$ for identical ones.  The mutual information is often described as the amount of information that the known labeling~$r$ tells us about the unknown one~$s$.  For our purposes, however, it is perhaps more useful to think of it as the amount of information we save if we already know~$r$ when specifying~$s$, as compared to if we don't.

In addition to increasing with the similarity of the labelings, the mutual information has the nice property of being symmetric in $r$ and~$s$.  In some circumstances, it is convenient to normalize it so as to create a measure whose value runs between zero and one~\cite{DDDA05}.  There are several ways to perform the normalization~\cite{MGH11}, but the most widely used normalizes by the mean of the entropies $H(r)$ and~$H(s)$, which preserves the symmetry with respect to $r$ and~$s$:
\begin{equation}
\mbox{Normalized mutual information} = {I(r;s)\over\half[H(r)+H(s)]}.
\label{eq:NMI}
\end{equation}

\subsection{Shortcomings of the mutual information}
\label{sec:problem}
As we have said, the mutual information is not perfect.  The particular shortcoming that we focus on here is illustrated by the following two simple examples.  First, consider a case where the labeling~$r$ consists of just a single group or community containing all $n$ objects.  At the same time let us assume that the labeling~$s$ is non-trivial, with $S>1$.  It is clear in this case that labeling~$r$ communicates no information about labeling~$s$, and hence the mutual information~$I(r;s)$ should be zero.  And indeed it is, since $P(r)=1$ and $P(r,s) = P(s)$ and hence
\begin{equation}
I(r;s) = \sum_s P(s) \log {P(s)\over P(s)} = 0.
\label{eq:example1}
\end{equation}

Now consider a second case where the labeling~$r$ consists of $n$ groups, each containing a single object.  It is again clearly true that $r$ communicates no information about~$s$ in this situation and hence the mutual information should be zero, but if we perform the calculation this is no longer what we find.  In this case $c_{rs}$ takes only the values zero and one and $a_r=1$ for all~$r$.  Combining Eqs.~\eqref{eq:sumrules1}, \eqref{eq:prs}, \eqref{eq:entropy}, and~\eqref{eq:mi}, we then get
\begin{align}
I(r;s) &= {1\over n} \sum_{rs} c_{rs} \log {n\,c_{rs}\over a_r b_s} \nonumber\\
  &=  {1\over n} \sum_{rs} \biggl[ c_{rs} \log c_{rs} - c_{rs} \log a_r
      - c_{rs} \log {b_s\over n} \biggr] \nonumber\\
  &= - \sum_s {b_s\over n} \log {b_s\over n}
   = - \sum_s P(s) \log P(s) \nonumber\\
  &= H(s),
\label{eq:example2}
\end{align}
(where, as previously, we are employing the convention that $0 \log 0 = 0$).

Equation~\eqref{eq:example2} is the complete opposite of what we expect.  The mutual information should be zero; instead it takes its maximal value of~$H(s)$.  The measure has failed completely.

\section{An improved measure of similarity}
\label{sec:improved}
In this paper we propose a new measure, the \defn{reduced mutual information}~$M$, which corrects for the failure illu\-strated in the previous section.  Our measure is simple to state---it is equal to the standard mutual information minus a single correction term, thus:
\begin{equation}
M = I(r;s) - {1\over n} \log \Omega(a,b),
\label{eq:M}
\end{equation}
where $\Omega(a,b)$ is a (usually large) integer equal to the number of $R\times S$ non-negative integer matrices with row sums~$a = \set{a_r}$ and column sums~$b = \set{b_s}$.  We give a (previously published) formula for calculating this number, to a good approximation, in Section~\ref{sec:tables}.

The origin of the correction term in~\eqref{eq:M} lies in the fact that even when our two labelings $r$ and $s$ agree perfectly, $r$~does not tell us everything about~$s$.  In order to deduce $s$ from $r$ one also needs to know how the values of the labels used in $r$ map to those used in~$s$.  One can encapsulate this mapping in a matrix and it is the information content of this matrix that gives rise to the term in~$\Omega(a,b)$.  To understand the argument in detail, let us look more closely at the conditional entropy and mutual information, paying careful attention to their derivation, including several terms that are commonly neglected.

\subsection{Information content of a labeling}
The central question we want to answer is this: if we know the labels in labeling~$r$, how much additional information is needed to specify the other labeling~$s$?  One can think in terms of two individuals, traditionally called Alice and Bob, who are in communication with one another.  Alice knows both labelings of the objects, $r$~and~$s$, but Bob only knows~$r$.  How much information, in the traditional Shannon sense, must Alice transmit in order to communicate $s$ to Bob?

The exact answer depends on how Alice encodes her communication with Bob.  There are more and less efficient ways to transmit the labeling~$s$ that will accomplish the desired outcome with shorter or longer communications.  At the simplest level, Alice could entirely ignore~$r$ and just send a complete record of $s$ to Bob as a single message.  Since there are $S$ possible labels for each object, there are~$S^n$ possible such messages Alice might need to send.  If the message is sent in binary, $k$~bits will suffice to encode uniquely each of these $S^n$ possibilities provided $2^k\ge S^n$, or $k \ge \log_2 S^n = n \log_2 S$.  Since $k$ is an integer, the number of bits needed to communicate $s$ in this manner is thus given by $\lceil n \log S \rceil$, the smallest integer not less than the log of the number of different messages one might send.  Then the amount of information \emph{per object}~is
\begin{equation}
H_1 = {1\over n} \lceil n \log S \rceil = \log S + \Ord(1/n).
\label{eq:H1}
\end{equation}
In the common case where the number~$n$ of objects is large we neglect the terms of order~$1/n$ and just say that the amount of information is~$\log S$.  (As before, the logarithm would traditionally be taken base~2, but the choice of base only affects the overall multiplier of the information and is not important for our purposes.  Henceforth, we will for convenience use natural logarithms in this paper, converting to units of bits only where necessary.)

\subsection{A more efficient encoding}
\label{sec:encoding}
Equation~\eqref{eq:H1} provides an upper bound on the amount of information Alice needs to send per object.  It is always possible to send a simple message like this and achieve the desired communication.  It is, however, often also possible to do better.  For example, suppose that not all labels occur with equal frequency, i.e.,~that $b_s$ varies with~$s$.  Then in most cases Alice can transmit~$s$ more compactly by first sending Bob the values of the~$b_s$, followed by the particular value of~$s$.  This approach works as follows.

\begin{figure}
\begin{center}
\includegraphics[width=8cm]{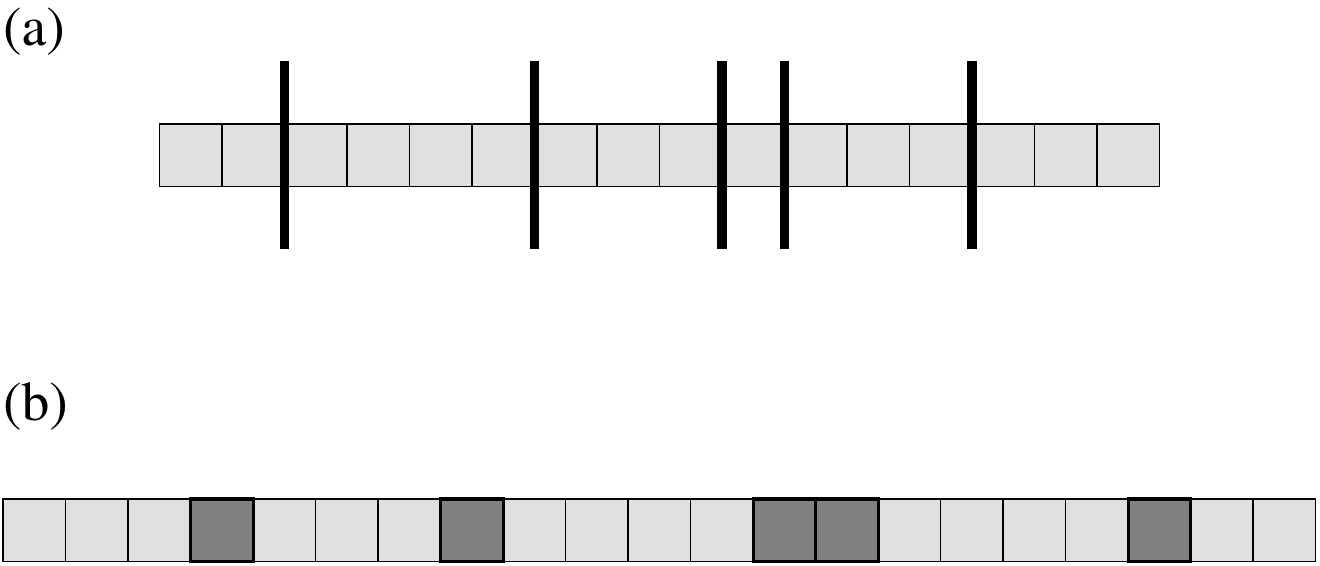}
\end{center}
\caption{(a)~The number of ways of choosing $S$ strictly positive integers $b_s$ such that they sum to $n$ is equal to the number of ways of placing $S-1$ dividers (the vertical bars) between $n$ objects (represented by the line of gray squares), with no two dividers allowed to fall in the same place.  The number of squares between each divider and the next represents the value of one of the~$b_s$, and trivially they sum to~$n$.  Since there are $n-1$ places the dividers can go, the total number of ways of choosing the~$b_s$ is ${n-1\choose S-1}$.  (b)~The number of ways of choosing $S$ integers~$b_s$ such that they sum to~$n$, where individual integers are now allowed to be zero (but not negative), is equal to the number of ways of placing an additional $S-1$ divider objects (the dark squares) among the $n$ lighter squares, for a total of $n+S-1$ squares overall.  The values of the $b_s$ are again the numbers of squares between the dividers, which again necessarily sum to~$n$.  Since two divider squares can now be adjacent, however, we can have $b_s=0$, and the total number of divisions is ${n+S-1\choose S-1}$.}
\label{fig:comb}
\end{figure}

The number of possible choices of~$b_s$ is equal to the number of different sets of $S$ non-negative integers that sum to~$n$, which is ${n-1\choose S-1}$, and the amount of information needed to communicate one of these sets is, once again, given by the log of this number.  (See Fig.~\ref{fig:comb} for a simple visual explanation of why this is the correct combinatoric factor.)  Once the values of the $b_s$ are specified, Alice need then only tell Bob which labeling $s$ is the correct one out of the set of all choices compatible with those values.  Normally this set of compatible choices is much smaller than the total number of possible labelings, which is why this approach is more economical.

The number of compatible choices of the labeling~$s$ is given by the multinomial coefficient $n!/\prod_s b_s!$ and the information needed to specify a particular choice is the logarithm of this number.  So the total information needed to communicate the value of~$s$ is $\log {n-1\choose S-1} + \log (n!/\prod_s b_s!)$, and the amount per object is
\begin{equation}
H_2 = {1\over n} \biggl[ \log {n-1\choose S-1} + \log {n!\over\prod_s b_s!}
                 \biggr].
\label{eq:H2}
\end{equation}

This expression is exact and easily computable.  We can, however, gain further insight by approximating the second term using Stirling's formula in the form
\begin{equation}
\log n! = n \log n - n + \half \log n + \Ord(1),
\label{eq:stirling}
\end{equation}
which gives us
\begin{align}
{1\over n} \log {n!\over\prod_s b_s!} &= \sum_s {b_s\over n} \log {n\over b_s}
  + \Ord\biggl({\log n\over n}\biggl) \nonumber\\
  &= - \sum_s P(s) \log P(s) + \Ord\biggl({\log n\over n}\biggl),
\label{eq:entropy2}
\end{align}
where as previously $P(s) = b_s/n$ is the probability that a randomly chosen object is in group~$s$.

The sum in Eq.~\eqref{eq:entropy2} is the Shannon entropy of~$s$, Eq.~\eqref{eq:entropy}.  Note that this entropy represents only the second term in Eq.~\eqref{eq:H2}.  It omits the first term, which measures the information needed to transmit the values of the~$b_s$.  Applying Stirling's approximation to this term too, we find that
\begin{equation}
{1\over n} \log {n-1\choose S-1} = (S-1) {\log n\over n} + \Ord(1/n),
\label{eq:vanishing}
\end{equation}
for $S$ constant as $n$ grows.  Thus this term is of the same order as the terms we have neglected in Eq.~\eqref{eq:entropy2} and it is often legitimate to neglect it too, in which case our information is given approximately by the conventional entropy, Eq.~\eqref{eq:entropy2}.

Assuming this to be the case, we can apply Jensen's inequality, which says that $\sum_s q_s \log x_s \le \log \sum_s q_s x_s$ for any set of non-negative numbers $x_s$ and weights~$q_s$ with $\sum_s q_s = 1$, with the exact equality applying only when all $x_s$ are equal.  Setting $q_s = b_s/n$ and $x_s = n/b_s$, we then get
\begin{equation}
H_2 = \sum_s {b_s\over n} \log {n\over b_s} < \log \sum_s 1 = \log S = H_1,
\label{eq:H2approx}
\end{equation}
except when all $b_s$ are equal.  Thus $H_2<H_1$ and this new scheme for transmitting the labeling~$s$ requires less information than simply sending the complete labeling.  In the language of information theory, we have found a more efficient encoding of the information by breaking it up and sending the values of $b_s$ and $s$ separately.

\subsection{Employing shared information}
\label{sec:shared}
So far Alice has made no use of the fact that Bob already knows the other labeling~$r$ of our $n$ objects.  If $r$ is correlated with $s$, however, then it may be possible to use that fact to convey the value of $s$ more succinctly.  Again there is more than one way to do this, corresponding to different encoding schemes, but we can illustrate the point with the following simple approach.  Alice first communicates to Bob the complete set of values~$c_{rs}$, defined in Section~\ref{sec:mi}, which are the numbers of objects with each combination of labels~$r,s$.  The matrix of values~$c_{rs}$ is known in this context as a \defn{contingency table}.  After sending the contingency table, Alice then follows up by sending the particular value of~$s$.  As in the previous section, knowing the contingency table narrows down the number of possible choices for $s$ and hence (in most cases) reduces the total amount of information Alice needs to send to specify the labeling~$s$.  The accounting works as follows.

Since both Alice and Bob already know~$r$ they also know~$a_r$, which are the row sums of the contingency table, so, even before Alice sends the table, Bob knows that it must have these row sums.  The number of choices for a row with $S$ elements which sum to~$a_r$ (and which are allowed to be zero) is ${a_r+S-1\choose S-1}$ (see Fig.~\ref{fig:comb} again for a visual explanation), and hence the number of tables with row sums~$a_r$ is $\prod_r {a_r+S-1\choose S-1}$ and the information needed to transmit the table is the log of this number.  At the same time, the number of possible choices of $s$ that are compatible with a particular contingency table is given by a product of multinomial coefficients $\prod_r \bigl[ a_r!/\prod_s c_{rs}! \bigr]$ and again the information is the logarithm.  Hence the total information, per object, needed to communicate~$s$ to Bob~is
\begin{equation}
H_3 = {1\over n} \sum_r \biggl[ \log\!{a_r+S-1\choose S-1}
      + \log {a_r!\over\prod_s c_{rs}!} \biggr].
\label{eq:H3}
\end{equation}

Focusing on the last term in this expression and again applying Stirling's approximation, Eq.~\eqref{eq:stirling}, we find that
\begin{align}
{1\over n} \sum_r \log {a_r!\over\prod_s c_{rs}!}
  &= \sum_{rs} {c_{rs}\over n} \log {a_r\over c_{rs}}
     + \Ord\biggl({\log n\over n}\biggl) \nonumber\\
  &= - \sum_{rs} P(r,s) \log {P(r,s)\over P(r)}
     + \Ord\biggl({\log n\over n}\biggl),
\label{eq:conditional2}
\end{align}
where $P(r,s)=c_{rs}/n$ is, as previously, the probability that a randomly chosen object has labels~$r,s$.  The sum in the second line of~\eqref{eq:conditional2} is precisely the conditional entropy~$H(s|r)$ of Eq.~\eqref{eq:conditional1} and thus we can to a good approximation write
\begin{equation}
H_3 \simeq H(s|r) + {1\over n} \sum_r \log\!{a_r+S-1\choose S-1}.
\label{eq:H3approx}
\end{equation}
Thus we see that the information needed to communicate~$s$ is not exactly equal to the standard conditional entropy, but contains an additional term.

Equation~\eqref{eq:H3approx} was proposed previously as a measure of the similarity of labelings by Dom~\cite{Dom02}, who argued for its use on the grounds that the second term compensates to some extent for the shortcomings of the conditional entropy discussed in Section~\ref{sec:problem}.

The measure has some issues however.  For instance, it does not equal~$H(s)$---as we would expect---in the case discussed in Section~\ref{sec:problem} where the labeling~$r$ places each object in a separate group on its own.  In that case, $H(s|r)=0$ and, putting $a_r=1$ for all~$r$, we get $H_3 = \log S$, which can be in error by a wide margin depending on the distribution of the~$b_s$.

Instead therefore we here adopt a different approach as follows.  In Section~\ref{sec:encoding} we found that specifying the values of the $b_s$ reduced the total amount of information Alice needed to send.  The same is true here.  Alice can use a three-part encoding, in which she first sends the values of the~$b_s$, then the contingency table, then $s$ itself.  Each step reduces the number of possibilities on the next and hence saves information.  The accounting is as follows.

First Alice sends the values of the~$b_s$.  As we have said, this takes an amount of information equal to $\log {n-1\choose S-1}$.  Second, she sends the contingency table.  Since she and Bob now know both $a_r$ and $b_s$ for all~$r,s$, both the row sums and the column sums of the table are fixed, and hence we need choose only among the set of tables that satisfy these constraints.  Suppose that the number of such tables is $\Omega(a,b)$.  Then the information needed to transmit the contingency table is~$\log \Omega(a,b)$.  Third, Alice sends $s$ itself, choosing among only those values that are compatible with the contingency table.  As before, this requires an amount of information equal to $\sum_r \log(a_r!/\prod_s c_{rs}!)$.  Thus the total information, per object, is
\begin{align}
H_4 = {1\over n} \biggl[ \log {n-1\choose S-1} + \log \Omega(a,b)
      + \sum_r \log {a_r!\over\prod_s c_{rs}!} \biggr].
\label{eq:H4}
\end{align}
If, as previously, the first term in this expression is negligible (since it vanishes for large~$n$---see Eq.~\eqref{eq:vanishing}), and given that the number~$\Omega(a,b)$ of contingency tables with row and column constraints can never exceed the number with row constraints alone, we necessarily have $H_4\le H_3$, making \eqref{eq:H4} the most efficient of the encoding schemes we have considered.  Technically, like all of the measures we have given, this one is still just an upper bound on the amount of information: in any particular situation there could exist an encoding that transmits~$s$ more succinctly.  However, as we argue in Section~\ref{sec:conclusions}, Eq.~\eqref{eq:H4} is the best bound we can establish without making use of domain-specific knowledge about the particular system of study.  If we wish to define a general measure that doesn't require additional tuning for each individual application, then Eq.~\eqref{eq:H4} is the best we can~do.

As discussed in Section~\ref{sec:mi}, it is in fact conventional to invert the scale on which information is measured by subtracting the information needed to communicate~$s$, given in this case by Eq.~\eqref{eq:H4}, from the amount needed when the receiver does not know~$r$, given by Eq.~\eqref{eq:H2}.  This gives us a mutual-information style measure thus:
\begin{align}
M &= H_2 - H_4 \nonumber\\
  & = {1\over n} \biggl[ \log {n!\over\prod_s b_s!}
  - \sum_r \log {a_r!\over\prod_s c_{rs}!} - \log \Omega(a,b) \biggr],
\label{eq:defsM}
\end{align}
which can be rearranged into the manifestly symmetric form
\begin{equation}
M = {1\over n} \biggl[ \log
   {n! \prod_{rs} c_{rs}!\over\prod_r a_r!\prod_s b_s!}
   - \log \Omega(a,b) \biggr].
\label{eq:mmi}
\end{equation}
We will call this quantity the \defn{reduced mutual information}.  It is equal to the amount of information that Alice saves by making use of the fact that Bob already knows~$r$.  Note that the first term in~\eqref{eq:H4}, which measures the amount of information needed to transmit the values of the~$b_s$, has canceled out, so in fact we need not assume (as we previously did) that this term is negligible.

Applying Stirling's approximation to the first term of~\eqref{eq:mmi}, we find that
\begin{equation}
{1\over n} \log {n! \prod_{rs} c_{rs}!\over\prod_r a_r!\prod_s b_s!}
  \simeq \sum_{rs} P(r,s) \log {P(r,s)\over P(r) P(s)},
\label{eq:hgapprox}
\end{equation}
which is the standard mutual information~$I(r;s)$ of Eq.~\eqref{eq:mi}, so, as claimed at the start of Section~\ref{sec:improved}, the reduced mutual information is, to a good approximation, equal to the standard mutual information minus the information needed to specify the contingency table $I(r;s) - (1/n) \log \Omega(a,b)$.  If one computes only the first term in this expression, as one normally does when comparing labelings, then one is neglecting the contingency table.  This is the root cause of the problem described in Section~\ref{sec:problem}.  For correct answers in all cases, one should compare labelings using the full expression, Eq.~\eqref{eq:mmi}, with both terms.

The reduced mutual information will normally be positive but it is possible for it to take negative values.  Referring to Eq.~\eqref{eq:defsM}, we see that this happens whenever $H_2<H_4$---in other words when it is more efficient for Alice to just send $s$ directly to Bob than to exploit correlations between $r$ and~$s$.  This tells us that in fact there \emph{are} no correlations, a least at the level where they give us useful information about the labeling, a result that could be useful in some situations.  The standard mutual information, by contrast, is never negative and hence has no natural threshold value that indicates absence of correlation between two variables.  The adjusted mutual information discussed in the introduction does have such a threshold, created by subtracting a small offset value from the mutual information~\cite{VEB10} but, as we argue in Section~\ref{sec:discussion}, Eq.~\eqref{eq:mmi} is more correct in most situations.

It is also possible to create a normalized version of Eq.~\eqref{eq:mmi}, akin to the normalized mutual information of Danon~\etal~\cite{DDDA05}, Eq.~\eqref{eq:NMI}, by dividing by the average of the values obtained when both labelings are equal to~$r$ and when both are equal to~$s$.  This gives a normalized information of
\begin{align}
&M_\textrm{norm} = \nonumber\\
  &\quad {\displaystyle
     2\log {n! \prod_{rs} c_{rs}!\over\prod_r a_r!\prod_s b_s!}
   - 2\log \Omega(a,b)\over\displaystyle
     \log {\rule{0pt}{9pt}n!\over\prod_r a_r!}
   + \log {n!\over\prod_s b_s!} - \log \Omega(a,a) - \log \Omega(b,b)},
\label{eq:normalized}
\end{align}
which is now equal to~1 when $r$ and $s$ are identical.

\subsection{Simple examples}
To illustrate the importance of the second term in~\eqref{eq:mmi} let us revisit our two simple examples from Section~\ref{sec:problem}.  In the first example, we considered a labeling~$r$ in which all objects belong to the same single group.  In this case the contingency table has just one row, and $S$ columns with entries~$n_{1s} = b_s$.  Hence all entries in the table are fixed and there is only one possible contingency table with the given row and column sums.  Thus the second term in~\eqref{eq:mmi} is zero and $M$ is given by the first term alone, which as we have seen is asymptotically equal to the standard mutual information.  In this case, therefore, we expect the standard measure to give the correct answer of zero, as indeed we found in Section~\ref{sec:problem}---see Eq.~\eqref{eq:example1}.

But now take our second example, in which the labeling~$r$ consists of $n$ groups of one object each.  In Section~\ref{sec:problem} we found that in this case the standard mutual information gave the wrong answer---it took on the maximal value of~$H(s)$ when common sense dictates that it should again be zero.  This is precisely because we neglected the information content of the contingency table.  In this case we have all $a_r=1$ and all $c_{rs}=0$ or~1, so that $a_r!=c_{rs}!=1$ for all $r,s$ and the first term in~\eqref{eq:mmi} is $\log(n!/\prod_s b_s!)$, which is indeed approximately equal to~$H(s)$ by Stirling's formula.  The second term, however, is now large.  The contingency table has $n$ rows and $S$ columns with a single~1 in each row and all other elements zero.  The number of such contingency tables that have the appropriate column sums~$b_s$ is equal to the number of ways of placing the $n$ objects in groups of sizes~$b_s$, which is given by the multinomial coefficient $n!/\prod_s b_s!$.  Hence the two terms in~\eqref{eq:mmi} cancel exactly to give $M=0$, which is the correct answer.

The intuitive explanation for this result is that in this case the contingency table in fact tells us the complete labeling~$s$.  Given that Bob knows the unique label~$r$ for each object, then once Alice tells him which value of~$s$ each $r$ corresponds to he can reconstruct $s$ without any further information.  Thus the contingency table contains as much information as $s$ itself, so Alice necessarily saves no information by sending the table first: she might as well have just sent $s$ directly.

\subsection{Counting contingency tables}
\label{sec:tables}
In order to compute the reduced mutual information of Eq.~\eqref{eq:mmi} we need to know the number~$\Omega(a,b)$ of possible $R\times S$ contingency tables with row and column sums equal to the given values of~$a_r$ and~$b_s$.  We are aware of no general closed-form expression for this number, although there are expressions for small (two- and three-row) tables~\cite{DG95}.  Tables can be counted numerically, for instance by exhaustively enumerating all $R\times S$ matrices with the desired row sums and counting how many have the desired column sums.  The number of matrices with row sums~$a_r$ is $\prod_r {a_r+S-1\choose S-1} \le {n+S-1\choose S-1}^R \le (n+S-1)^{RS}$ and hence the running time for this procedure is polynomial in~$n$, but in practice the degree of the polynomial is so large as to make the method unworkable for all but the smallest of matrices.  More sophisticated algorithms have been proposed with somewhat better running times~\cite{DG95,Greselin03}, but they are still practical only for small cases.  For larger values of $n$, $R$, and~$S$ one can calculate approximate answers by Monte Carlo sampling~\cite{JVV86,DG95}, but a more convenient solution for our purposes is to make use of analytic approximations such as those given in Refs.~\cite{BBK72,Bender74,DG95,CM07b,Barvinok12}.  We are particularly interested in two limits.  The first is the sparse limit, typified by our example above in which each object is placed in a group on its own.  In cases like this where most elements of the contingency table are zero and the elements that are nonzero are small, B\'ek\'essy~\etal~\cite{BBK72,DG95} have shown that
\begin{align}
\log \Omega(a,b) \simeq \log {n!\over\prod_r a_r! \prod_s b_s!}
  + {2\over n^2} \sum_r {a_r\choose 2} \sum_s {b_s\choose 2}.
\label{eq:bbk}
\end{align}
Applying Stirling's formula we then have
\begin{equation}
{1\over n} \log \Omega(a,b) \simeq H(s) - {1\over n} \sum_r \log a_r!
  + \Ord(1/n).
\end{equation}
Thus in general the contingency table requires an amount of information on the order of~$H(s)$ per object for its specification, making it essential that it be included in the calculation if we are to obtain useful results. 

Substituting Eq.~\eqref{eq:bbk} into Eq.~\eqref{eq:mmi} we get
\begin{equation}
M \simeq {1\over n} \sum_{rs} \log c_{rs}!
  - {2\over n^3} \sum_r {a_r\choose 2} \sum_s {b_s\choose 2},
\end{equation}
which can be used to calculate~$M$ in this regime.

This, however, is not the regime in which we are most commonly operating.  Usually we are interested in cases where the numbers of groups $R$ and $S$ are substantially smaller than~$n$ and the contingency table is relatively dense, with many elements much larger than~1.  In this regime we use a different approximation, a symmetrized version of a formula due to Diaconis and Efron~\cite{DE85}:
\begin{align}
&\log \Omega(a,b) \simeq (R-1)(S-1) \log(n+\half RS) \nonumber\\
  &\qquad{} + \half(R+\nu-2) \sum_s \log y_s
            + \half(S+\mu-2) \sum_r \log x_r \nonumber\\
  &\qquad{} + \half \log {\Gamma(\mu R) \Gamma(\nu S)\over
              [\Gamma(\nu)\Gamma(R)]^S [\Gamma(\mu)\Gamma(S)]^R}\,,
\label{eq:diaconis}
\end{align}
where
\begin{align}
w   &= {n\over n+\half RS}, \\
x_r &= {1-w\over R} + {w a_r\over n}, \qquad
y_s  = {1-w\over S} + {w b_s\over n}, \\
\mu &= {R+1\over R\sum_r y_r^2} - {1\over R}, \hspace{2.18em}
\nu  = {S+1\over S\sum_r x_r^2} - {1\over S}.
\end{align}
The leading term in Eq.~\eqref{eq:diaconis} is of order $RS\log n$ and hence
\begin{equation}
{1\over n} \log \Omega(a,b) = \Ord\biggl({RS\log n\over n}\biggr).
\label{eq:orderofmag}
\end{equation}
If $R$ and $S$ are asymptotically constant as~$n$ becomes large this expression is of the same order of magnitude as terms neglected in the conventional definition of the mutual information, Eq.~\eqref{eq:conditional2}, and therefore can itself be neglected for large enough~$n$.  This is why the traditional mutual information works well in some circumstances, but there are also common situations in which these conditions do not apply.  First, either or both of the number of groups $R$ and $S$ may increase with~$n$.  For instance, in community structure problems in networks it is reasonable to suppose that $R$ and/or $S$ would grow: the number of groups of friends in the world, for instance, probably grows with world population.  In this case~\eqref{eq:diaconis} cannot be neglected for large~$n$.  Indeed if $R$ and $S$ are of order~$\sqrt{n}$ or larger, which they often are, then~\eqref{eq:diaconis} cannot be neglected under any circumstances.

Second, in practice $n$ is not infinite and in many cases it is not even particularly large: values of a few dozen to a few hundred are common in practical classification problems.  In this regime~\eqref{eq:diaconis} may well be comparable with the standard mutual information and again must be included to get accurate results.

\begin{figure*}
\begin{center}
\hfill\includegraphics[width=7cm]{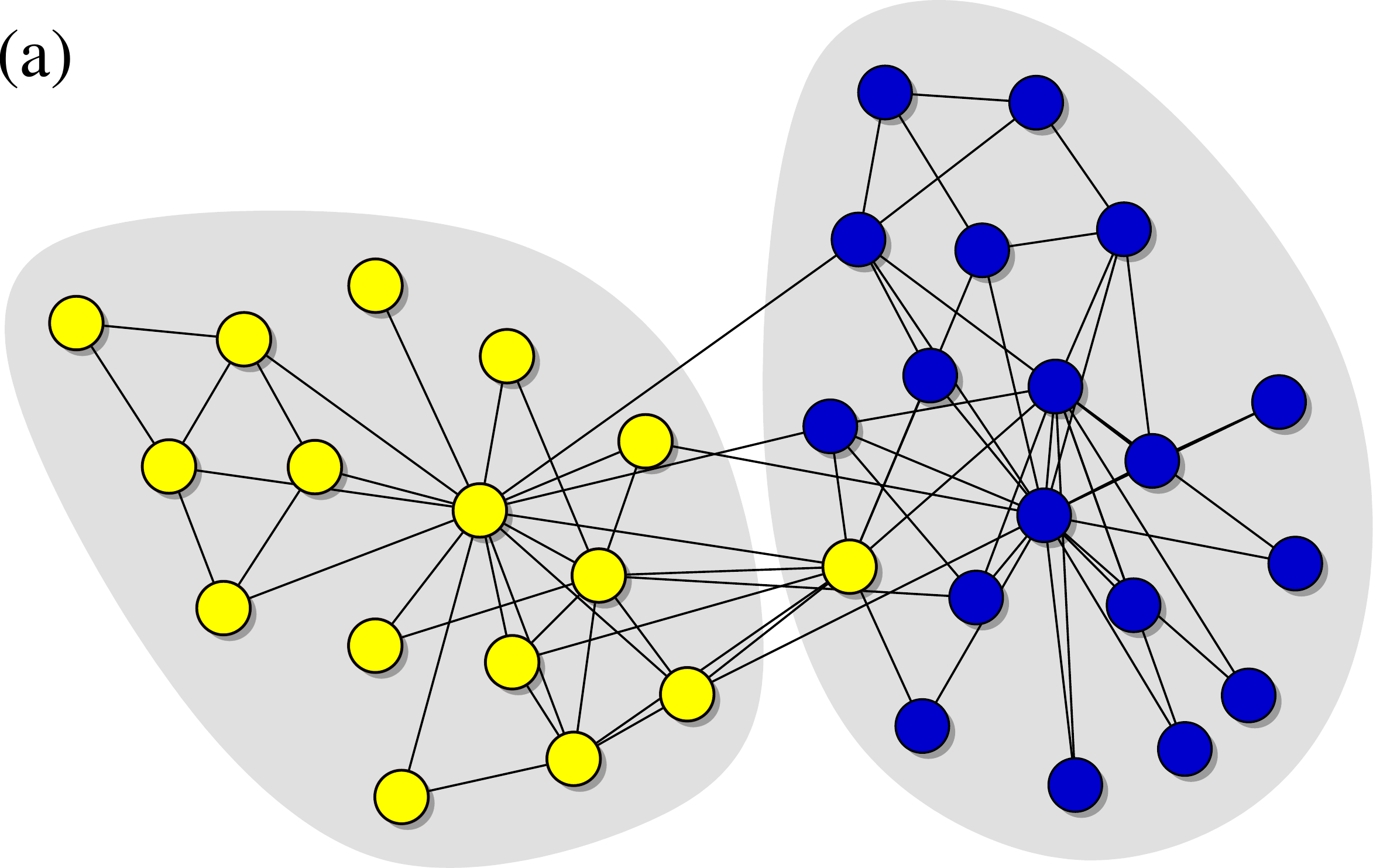}\hfill
\includegraphics[width=7cm]{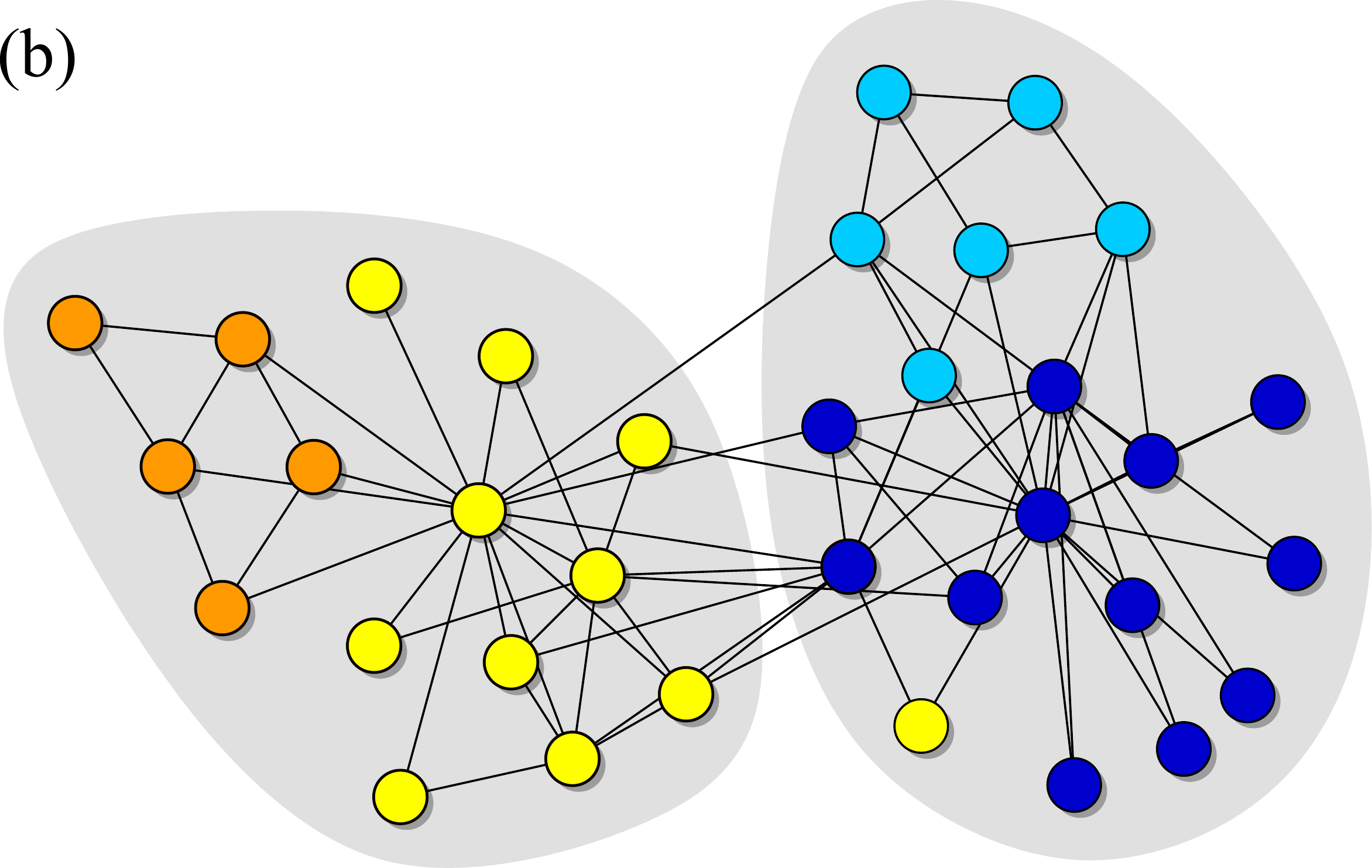}\hfill\null
\end{center}
\caption{Two divisions into communities of the ``karate club'' network of Zachary~\cite{Zachary77}.  (a)~Division found by statistical inference, maximizing the profile likelihood for the degree-corrected stochastic block model~\cite{KN11a}.  (b)~Division found by maximizing modularity~\cite{Newman06b}.  The accepted ground-truth division is indicated by the shaded regions.  According to the standard mutual information, division~(b) has the stronger similarity to the ground truth, but division~(a) is more similar according to the reduced mutual information of Eq.~\eqref{eq:mmi}.}
\label{fig:zachary}
\end{figure*}

\section{Example applications}
\label{sec:applications}
As a simple example of the use of our measure, consider Fig.~\ref{fig:zachary}, which depicts community structure in a well-known test network, the ``karate club'' network of Zachary~\cite{Zachary77}.  This network, which depicts friendships among a group of university students, is widely agreed to divide into two communities, represented by the shaded areas in the figure.  Panels (a) and~(b) in the figure show two different possible divisions of the network found using popular community detection methods, the first by a statistical inference technique~\cite{KN11a} and the second by the method of maximum modularity~\cite{Newman06b}.  The two divisions are quite different.  The first is closely similar to the accepted ground truth, but differs from it for one node in the center of the figure.  The second, by contrast, divides the network into four smaller communities, which arguably align quite well with the ground truth, although again one node is clearly wrong.  Which of these two divisions is more similar to the ground truth?  Most observers would probably say the two-group division on the left.  Both get a single node wrong, but the two-group division is closer in overall form to the ground truth.

Traditional mutual information, however, says the reverse.  Mutual information scores are 0.788 bits per node for the two-group division and 0.807 for the four-group division, indicating that the four-group division is slightly superior.  If, however, one includes the information in the contingency table the outcome is reversed.  For this small example we have no need of the approximation in Eq.~\eqref{eq:diaconis} to calculate the number of contingency tables---the tables can be counted by exhaustive enumeration and we find that $\Omega(a,b) = 16$ in the first case and 428 in the second.  Substituting these figures into Eq.~\eqref{eq:mmi} we find a reduced mutual information of 0.670 bits per node for the two-group division and 0.550 for the four-group one, making the two-group division now the favored choice.

\begin{figure*}
\begin{center}
\includegraphics[width=16.5cm]{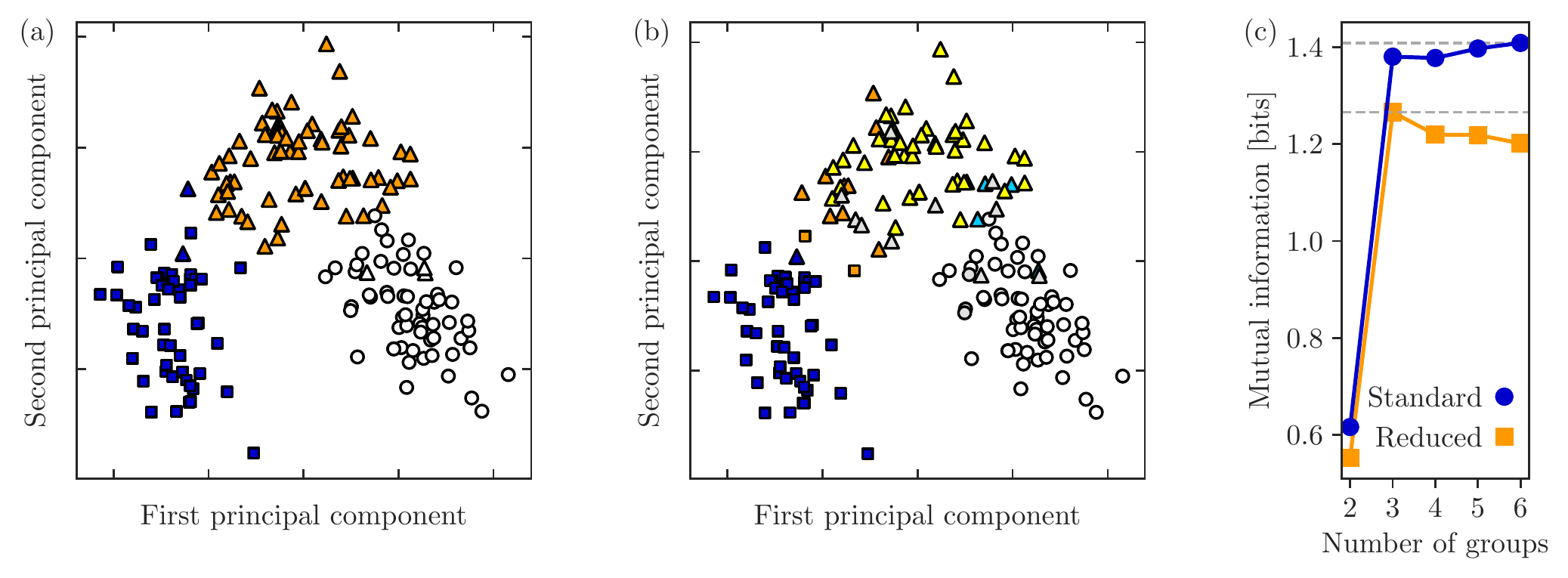}
\end{center}
\caption{Clustering analysis of the wine data set described in the text~\cite{BM98}, which represents 12 measured attributes of 178 different Italian wines.  (a)~A three-group clustering of the data found using $k$-means applied to a normalized version of the data.  The data points are projected onto the data set's two most significant principal components, with colors representing the calculated clustering and shapes representing the ground truth.  (b)~A six-group clustering of the same data, again computed using $k$-means.  (c)~The standard mutual information and the reduced mutual information of Eq.~\eqref{eq:mmi} for divisions into 2, 3, 4, 5, and~6 groups using $k$-means.  While the standard measure increases with increasing number of groups, the modified measure peaks at the correct ground-truth value of three.}
\label{fig:wine}
\end{figure*}

A more complex example is depicted in Fig.~\ref{fig:wine}, which shows results for a widely studied data set from the machine learning literature representing 178 different wines, each produced in the same region of Italy and each made from one of three different grapes~\cite{BM98}.  Chemical and photometric analysis of the wines produces 13 different measurements for each one, for attributes such as alcohol content, flavinoids, and color intensity.  One of these attributes we discard based on standard feature selection criteria~\cite{Pedregosa11}.  The remaining 12 can be used to place each wine at a point in a 12-dimensional data space, and previous studies have shown that it is possible, by clustering these points, to infer at least approximately which wine was made from which grape.  Figure~\ref{fig:wine} shows the results of clustering using what is perhaps the simplest of methods, the $k$-means algorithm.  Panels~(a) and~(b) show two such clusterings, one into three groups (which is the ``correct'' number according to the ground truth) and the other into six.  In both panels the calculated clustering is indicated by the colors of the data points while their shapes indicate the ground truth, and most observers would probably say that the division into three groups in panel~(a) is closer to the ground truth---remarkably it classifies all but five of the wines correctly.

Once again, however, traditional mutual information reaches the opposite conclusion, saying that the division into six groups is better: the three-group division has a mutual information of 1.380 bits per data point, while the six-group one has a slightly better score of 1.409.  If one includes the correction due to the information content of the contingency table, however, the results are reversed.  In this case it is impractical to count the number of contingency tables exhaustively: for the six-group division there are estimated to be over $10^{11}$ tables.  Instead, therefore, we make use of the approximate formula in Eq.~\eqref{eq:diaconis}.  Combining this formula with Eq.~\eqref{eq:mmi} we find a reduced mutual information of 1.266 bits per data point for the three-group division and a lower 1.202 for the six-group one, so that the three-group division is now favored.  Figure~\ref{fig:wine}c shows results for both the standard and reduced mutual information for $k$-means clusterings of the data points into 2, 3, 4, 5 and~6 groups.  As the figure shows, the standard measure increases with the number of groups while the modified measure peaks at the correct value of three then falls off.

\section{Comparison with other measures}
\label{sec:discussion}
As discussed in the introduction, the measure we give for comparing clusterings is not the only one proposed for this task.  In this section we review some of the previously proposed measures and their relationship to our reduced mutual information.

In addition to the standard mutual information, whose disadvantages we have already discussed in depth, the  measures that are most directly related to ours are the measure proposed by Dom in~\cite{Dom02}, the adjusted mutual information proposed by Vinh~\etal~\cite{VEB10}, and the variation of information proposed by Meil\u{a}~\cite{Meila07}.  Let us consider each of these in turn.

The measure proposed by Dom~\cite{Dom02} is equal to our measure~$H_3$, Eq.~\eqref{eq:H3}, when the conditional entropy term is approximated using Stirling's formula as in Eq.~\eqref{eq:H3approx}.  This measure does compensate to some extent for the shortcomings of the standard conditional entropy but, as discussed Section~\ref{sec:shared}, it also has some undesirable features.  For instance, it does not take the expected value of~$H(s)$ when labeling~$r$ places every node in a separate group of size one, and it is moreover normally larger than our measure~$H_4$ and hence necessarily a poorer approximation to the amount of information needed to specify~$s$---both measures by definition give upper bounds on the information, but $H_4$ gives a lower, and hence better, bound.  In that the stated goal of both measures is to calculate the information, it is therefore normally better to use~$H_4$ or, equivalently, the reduced mutual information~$M$.

A more subtle case is presented by the adjusted mutual information~\cite{VEB10} (and variants such as the relative mutual information of Zhang~\cite{Zhang15}).  Like our measure, the adjusted mutual information subtracts a correction term from the standard mutual information~$I(r;s)$.  For the adjusted mutual information this term is equal to the average of $I(r;s)$ over all possible choices of the labels~$s$.  It is helpful in comparing this measure to our own to re-express this adjustment in terms of counts of contingency tables as follows.

As Vinh~\etal\ point out, if all labelings~$s$ are equally likely then all contingency tables are not.  For a given labeling~$r$, the same table can be generated by many different choices of~$s$.  As discussed in Section~\ref{sec:shared}, the number of labelings~$s$ that correspond to a particular contingency table is $\prod_r a_r!/\prod_{rs} c_{rs}!$, so if all $s$ are equally likely then the probability~$Q_T$ of finding a particular table~$T$ is equal to this number divided by the total number of labelings~$s$, which is $n!/\prod_s b_s!$.  Thus,
\begin{equation}
Q_T = {\prod_r a_r!/\prod_{rs} c_{rs}!\over n!/\prod_s b_s!}
    = {\prod_r a_r! \prod_s b_s!\over n! \prod_{rs} c_{rs}!}.
\label{eq:hypergeometric}
\end{equation}
This probability distribution over contingency tables is known as the multivariate hypergeometric distribution.

Given this distribution, information theory tells us that the amount of information needed to transmit a particular table~$T$ is
\begin{equation}
- \log Q_T = \log {n! \prod_{rs} c_{rs}!\over\prod_r a_r! \prod_s b_s!}
  \simeq n I(r;s),
\end{equation}
where we have used Eq.~\eqref{eq:hgapprox}.  Averaging over all tables~$T$, we then find that the average information per object needed to transmit the contingency table is
\begin{equation}
- {1\over n} \sum_T Q_T \log Q_T = \bigl\langle I(r;s) \bigl\rangle,
\label{eq:amiinfo}
\end{equation}
where $\langle\ldots\rangle$ denotes the average over the hypergeometric distribution.  This is precisely the average mutual information that Vinh~\etal\ subtract in order to create their adjusted mutual information.

In other words, though this is not how it was originally motivated, the adjusted mutual information can be interpreted in a manner analogous to our own measure, as the standard mutual information minus the amount of information needed to specify the contingency table.  The crucial difference, however, is that in calculating the correction Vinh~\etal\ assume that all \emph{labelings~s} are equally likely.  This contrasts with our own measure, Eq.~\eqref{eq:mmi}, which assumes that all \emph{contingency tables} are equally likely.

We argue that the latter is more correct in the sense of giving a more accurate estimate of the information needed to transmit~$s$ in realistic situations.  The probability distribution~$Q_T$ over contingency tables in Eq.~\eqref{eq:hypergeometric} is strongly non-uniform, which has a large effect on the information calculated in Eq.~\eqref{eq:amiinfo}, but it is non-uniform in an unrealistic way.  If we assume that all labelings~$s$ are equally likely then we expect the elements of the rows of the contingency table to be roughly equal: there are vastly more choices of~$s$ that give roughly equal elements than significantly unequal ones, and hence $Q_T$ is weighted heavily in favor of such tables.  But if the elements of the table are equal then $r$ and $s$ are uncorrelated, which defeats the point of calculating the mutual information in the first place.  We are, by definition, interested in mutual information measures in situations where $r$ and $s$ are correlated a substantial fraction of the time.  And if they are correlated then the table will have significantly unequal elements and Eq.~\eqref{eq:amiinfo} will be a poor estimate of the amount of information needed to transmit it.  Indeed in some situations one can show that Eq.~\eqref{eq:amiinfo} is off by a factor as large as $\Ord(n/\log n)$, making it very inaccurate indeed.  Our own measure, by contrast, assumes a uniform distribution over contingency tables, making it unbiased with respect to the form of the table.  On a purely practical level, the adjusted mutual information is also quite time-consuming to calculate, whereas our own measure can be calculated in time $\Ord(RS)$ when the contingency table is known or $\Ord(n)$ when it is not, which is the same time complexity as the standard mutual information.

Third, let us consider the variation of information~\cite{Meila07}, which is a somewhat different animal.  It is defined as
\begin{equation}
\mbox{Variation of information} = H(s|r) + H(r|s).
\end{equation}
Unlike the mutual information (and similar measures) the variation of information is a \emph{dissimilarity} measure, taking larger values for unlike labelings.  The variation of information is satisfactory in that it does not suffer from the failings of the simple mutual information.  When labeling~$r$ places every object in a group of its own, for instance, it gives an answer that diverges as $n$ becomes large.  Arguably this is not exactly the answer we are looking for, but it is definitely nonzero, indicating that the two labelings are very different from one another.  Moreover the variation of information has a number of attractive formal properties.  In particular, unlike any of the other measures we have considered (including our own), it is a true metric distance measure: it is symmetric, non-negative, and satisfies the triangle inequality (meaning that the distance, in terms of variation of information, from $r$ to $t$ is never greater than the sum of the distance from $r$ to $s$ and from $s$ to~$t$).  This means one can unambiguously say when one pair of labelings are closer together than another and hence define a topology on the space of all labelings.

The main disadvantage of the variation of information is that it does not have a simple interpretation in terms of information content.  It is not, for example, equal to the information needed to transmit the unknown labeling~$s$ under any encoding we are aware of.  Instead it is equal to the information needed to transmit~$s$ given~$r$ plus the information needed to transmit $r$ given~$s$, not including the information needed to send the contingency table.  This makes it a less natural choice for the type of applications we are considering.

\section{Conclusions}
\label{sec:conclusions}
Mutual information is widely used as a quantitative measure for comparing competing labelings or divisions of a set of objects, such as arise in classification problems in machine learning or community detection problems in network science.  In this paper we have argued, however, that the standard mutual information is not always adequate for this purpose, giving incorrect answers---sometimes maximally incorrect---particularly when the labelings have different numbers groups.  We have shown how this shortcoming can be rectified by the inclusion of an additional term in the mutual information equal to minus the logarithm of the number of contingency tables compatible with the sizes of the groups.  We have given two example applications of our proposed measure, showing how the inclusion of the additional term can make not only quantitative but also qualitative differences to the outcomes of real-world calculations.  We have also given a detailed comparison of our measure to other variants of the mutual information and argued that in practical situations it better captures the similarity of pairs of labelings than its competitors.

In closing, let us consider whether it would be possible to further improve our measure beyond what is described here.  As we have said, all measures of the information needed to describe a labeling are actually upper bounds on the true information.  In any individual case it may be possible that a labeling has some special form or structure that makes it possible to communicate it more succinctly.  If one could calculate the true minimum amount of information needed to transmit a labeling in all cases then by definition this would give a better measure than any of those discussed in this paper.

There are, however, a number of reasons why this point of view is not a helpful one.  First, calculating the true minimum information is normally impossible.  It would be equivalent to calculating of the so-called Kolmogorov complexity~\cite{WD99}, whose computation is provably impossible in general.  Second, a measure such as our reduced mutual information, Eq.~\eqref{eq:mmi}, has the desirable feature of expressing the information as a correction to the conventional mutual information.  In effect, it decomposes the information into the standard message that one would send under traditional information theory plus an additional term that has previously been neglected.  Thus it gives us not only a better measure of similarity but it tells us in a straightforward manner how that measure is related to others that we are familiar with.

Even if we focus on measures with this latter property, however, and if we abandon the hope of finding the true minimum information, it is still possible that we could find a better bound.  Given that we are fixing the mutual information term, such an improvement would have to come from finding a more efficient scheme for transmitting the contingency table (or some equivalent object).  As discussed in Section~\ref{sec:discussion}, we have assumed that all contingency tables are equally likely, which by definition maximizes the information needed to send them.  If, in fact, some tables are more likely than others, then we can in principle use that fact to improve our encoding and transmit the table more succinctly.  The adjusted mutual information, discussed in Section~\ref{sec:discussion}, offers one way of doing this, but the particular probability distribution over tables that it employs is, we have argued, incorrect.  If we could, however, accurately characterize the true distribution of tables in any particular system, then we could in principle find a better encoding for labelings.  This means, unfortunately, that any improved information measure would have to make use of domain-specific knowledge about the particular application of interest.  In the absence of any such knowledge, the uniform distribution over contingency tables gives the only bound on the information that is correct in all cases and hence we argue that the measure proposed in this paper is the natural choice for classification tasks.

\begin{acknowledgments}
This work was funded in part by the US National Science Foundation under grant DMS--1710848 (MEJN) and by the James S. McDonnell Foundation (JGY).
\end{acknowledgments}

\end{document}